\documentclass[twocolumn,showpacs,preprintnumbers,amsmath,amssymb]{revtex4}

\usepackage{epsfig}
 
\usepackage{graphicx}% Include figure files
\usepackage{dcolumn}% Align table columns on decimal point
\usepackage{bm}% bold math

\begin{document}

\preprint{1.0/03}

\def\be {\begin {equation}}
\def\ee {\end {equation}}
\def\beq {\begin {eqnarray}}
\def\eeq {\end {eqnarray}}

\title{The isotope effect in impure high $T_c$ superconductors.}
\author{M. Mierzy\'nska}
\author{K. I. Wysoki\'nski}
\affiliation{Institute of Physics, M. Curie-Sk\l odowska University,\\ 
Radziszewskiego 10, Pl 20-031 Lublin, Poland} 
 
\date{\today}

\begin{abstract}
 The influence of various kinds of impurities on
the isotope shift exponent $\alpha$ of  high temperature superconductors
has been studied. In these materials
the dopant impurities, like $Sr$ in $La_{2-x}Sr_xCuO_4$,
play different role
and usually occupy different sites than impurities
like Zn, Fe, Ni {\it etc} intentionally introduced into the system to 
study  its superconducting properties.
 In the paper the in-plane and out-of-plane impurities
present in layered superconductors have been
considered. They  differently affect the 
superconducting transition temperature $T_c$.
The relative change of isotope shift  coefficient,
however, is an universal
function of  $T_c/T_{c0}$ ($T_{c0}$ reffers to 
impurity free system)
 {\it i.e.} for angle independent scattering rate and density of states 
 function it does not depend whether
the change of $T_c$ is due to in- or out-of-plane impurities. 
The role of the
anisotropic impurity scattering in changing
oxygen isotope coefficient
of superconductors with various symmetries of the order
parameter is elucidated. The comparison of the calculated  
and experimental
dependence of $\alpha/\alpha_0$,
where $\alpha_0$ is the clean system
isotope shift coefficient, on $T_c/T_{c0}$
is presented for a number of cases studied. 
The changes of $\alpha$ calculated within
stripe model of superconductivity
in copper oxides resonably well describe the
data on $La_{1.8}Sr_{0.2}Cu_{1-x}(Fe,Ni)_xO_4$, without 
any fitting parameters.
\end {abstract}
%\pacs{  74.25.Bt Thermodynamic properties
%	     74.62.-c Transition temperature variations
%     74.62.Dh Effects of crystal defects, 
%    doping and substitution}

\maketitle

\section {Introduction}

It is hard to overestimate the role played by the impurities 
introduced into otherwise clean superconductor. In response 
to impurities the  superconducting properties of the material
change. The changes include superconducting transition
temperature, slope and jump of the specific heat, upper 
critical field, superfluid density, and other thermodynamic 
and electromagnetic characteristics \cite{Genn,Tink}.

 One of the parameters of great experimental and theoretical
importance is the
isotope coefficient $\alpha$ defined by the power law
dependence of superconducting
transition temperature $T_c$ on the isotope mass $M$ of
 the element: $T_c \propto M^{-\alpha}$
or $\alpha=-\frac{\partial \ln{T_c}}{\partial \ln{M}}$.
Illegal variable name.
In the BCS model of
superconductivity it has been predicted to take on universal
value $\alpha_{BCS} =\frac{1}{2}$
and verified experimentally for a number of  superconducting 
elements and simple compounds.
 In chemically complex multicomponent systems one usually 
 defines the partial coefficients
$\alpha_{i} = - {\partial \ln{T_c} \over \partial \ln{M_i}}$,
where $M_i$ is the isotope mass of the i-th component.
In high temperature superconductors (HTS) typically
  $O^{18}$ is replaced \cite{Bat,Lea,Bou} by $O^{16}$.  This 
defines so called oxygen isotope coefficient $\alpha^{O}$.  Limited
data are available on the copper isotope shift $\alpha^{Cu}$ in
these materials \cite{Fra1,Fra2}.

The effect of impurities introduced into the superconductor 
on its $T_c$ strongly depends on the symmetry of the order 
parameter. It is known that non magnetic
impurities hardly change
$T_c$ of s-wave superconductors (Anderson theorem).
On the other hand non magnetic impurities
are effective pair breakers in spin singlet d-wave
and spin-triplet p- or f- wave superconductors.
On the contrary magnetic impurities break time reversal
symmetry and strongly affect $T_c$ of all
superconductors including s-wave ones. Changing
$T_c$ of the material, impurities indirectly
influence all its parameters. In particular this is
true for isotope coefficient $\alpha$. If one finds
the change of the superconducting transition
temperature due to impurities as
${T_c \over T_{c0}}=f(T_{c0},\gamma)$, where $\gamma$
symbolizes all
relevant parameters other than $T_{c0}$ itself, than
${\alpha \over \alpha_{0}}=
{\partial \ln{T_c} \over \partial \ln{M}}/{\partial 
\ln{T_{c0}} \over \partial \ln{M}}$ can easily be 
calculated from the explicit knowledge of function 
$f(T_{c0},\gamma)$.

It is the purpose of this work to systematically study
 the effect of disorder on the isotope
effect of superconductors by {\it assuming} that 
dependence of ${\alpha \over \alpha_o}$ on
${T_c \over T_{co}}$ can be solely attributed to the
effect of impurities.
 The impurities
introduced into the superconductor  modify
quasi-particle spectrum, interaction parameters
and induce pair breaking.
This results in a change of the superconducting 
transition temperature and the isotope coefficient.

The motivation for the present analysis
partially comes from the recent experiments which
suggest strong effect of electron-phonon coupling
on the dynamics of electrons in high-temperature
superconductors \cite{Lan,Shen}. 
 
The shift of $T_c$ with ionic mass was a crucial experiment 
to confirm the electron-phonon mechanism of superconductivity 
in BCS superconductors. Similarly the systematic studies of 
various isotopic substitutions in HTS are important to
understand the role of electron-phonon interaction in 
these materials. After early experiments with contradictory 
conclusions \cite{Bat,Lea,Bou,Ron} 
it has later been unequivocally established that the oxygen 
isotope shift $\alpha$ is non-zero, and takes smallest value 
for optimally doped materials. It increases when one moves 
into underdoped region.

Our paper  extends
 the recent work of Openov et al. \cite{openov}, who consider 
 the effect of magnetic and non-magnetic impurities in s-wave
and d-wave superconductors and that of Kresin and coworkers 
\cite{Kresin} who study the isotope effects in s-wave 
superconductors doped with 
 magnetic impurities and those showing the dynamic Jahn-Teller and proximity 
effects.  Our aim here is to elucidate the role (in changing $\alpha$)
  of the out-of-plane impurities 
in layered systems and the effect of impurity anisotropy.
 We also analyze the change of isotope effect due to $Zn$ 
 impurities in the striped phase model of high $T_c$ 
 superconductors. In view of the recent interest in superconductivity
and the critical role of impurities played in
$Sr_2RuO_4$ possibly spin-triplet superconductor \cite{Rice} 
we shall consider isotope coefficient for p-wave
order parameter. 

In section 2 we explain the approach
 and apply it to the
layered systems in which carriers 
%responsible for normal
%transport and superconductivity 
mainly reside in
active  layers ($CuO$ in HTS), while impurities are placed
 either in the active or in the passive
layer. It turns out that due to low angle scattering
  these out-of-plane impurities have
small effect on superconducting transition temperature.
 However their effect on the isotope
coefficient is universal in a sense that $\alpha/\alpha_0$
depends only on $T_c/T_{c0}$. 
In section 3 we analyze the role of in plane anisotropic 
scatterers in changing transition temperature and the 
isotope coefficient. Section 4 
contains discussion of the results and
comparison with experimental data.
The predictions of the change
of $T_c$ by $Zn$ impurities obtained recently by the stripe
theory superconductivity lead to
the isotope coefficient, which is calculated in section 5.
We end up with conclusions.

\section{ In-plane and out-of-plane impurities in
layered  superconductors}

In this section we shall compare the effect of in-plane
and out-of-plane impurities in high temperature
superconductors on $T_c$ and $\alpha$. The Hamiltonian of 
the superconductor containing both kinds of impurities 
takes the form
\beq
H & = & \sum_{\vec k\sigma} (\varepsilon_{\vec k} - \mu)
c^+_{\vec k\sigma} c_{\vec k\sigma} + \sum_{kk'\sigma} V_{\rm in}
(\vec k, \vec k')  c^+_{\vec k\sigma} c_{\vec k'\sigma} \\
  &+& \sum_{\vec k\vec k'\sigma} V_{\rm out} (\vec k,\vec k') c^+_{\vec
k\sigma} c_{\vec k'\sigma} + \sum_{\vec k\vec q} V_{\vec k,\vec q}
c^+_{\vec k\uparrow} c^+_{-\vec k\downarrow} c_{\vec q\downarrow}
c_{\vec q\uparrow}\,,
\nonumber
\eeq
where $\varepsilon_{\vec k}$ is the single-particle energy, 
$\mu$ --chemical potential $c^+_{\vec k\sigma} (c_{\vec k\sigma})$
 denotes creation (annihilation) operator for an spin $\sigma$ electron
 in a state $\vec k$.
 The second term describes the 
scattering of carriers by in-plane while third by out-of-plane 
impurities. The pair potential $V_{\vec k\vec q}$ is assumed to 
take on the separable form 
$V_{\vec k\vec q} = -V_0\phi(\hat k) \phi(\hat q)$ dependent 
on the wave vector direction $\hat k = {\vec k\over |\vec k|}$. 
The superconducting order parameter is defined by
\be
\Delta(\vec k) = \sum_{\vec k} V_{\vec k \vec q} \langle c_{-\vec q\downarrow}
c_{\vec q\uparrow}\rangle
\ee
and the selfconsistent equation for it, easily derived by 
the Green's function technique, reads \cite{abrikgor}
\be
 \Delta(\vec k) = \sum_q V_{kq} {1\over\beta} \sum_{\omega_n}
{\tilde\Delta(q)\over\tilde\omega^2_n + (\varepsilon_q - \mu)^2 +
|\tilde \Delta(\vec q)|^2},
\label{gap}
\ee
where $\tilde\omega_n $ and $\tilde\Delta(q)$ are
renormalised frequency and order parameter in impure system.

The correction to the self-energy due to impurity scattering
$ \Sigma_k (i\omega_n)=G^{-1}_{0k} - G^{-1}_{k} $,
where $\omega_n = (2n+1)\pi k_B T_c$ is the Matsubara frequency,
$G_{0k}$ and $G_{k}$ are the Nambu - Gorkov
 (imaginary time) Green`s functions,  is calculated here
in the Born approximation \cite{abrikgor,Maki}
\beq
 \Sigma(\vec k,i\tilde{\omega}_n)  &=&  n_{\rm in} \sum_{\vec q}
| V_{\rm in}(\vec k - \vec
 q)|^2 \hat \tau_3 G_q (i\tilde{\omega}_n)\hat \tau_3
\nonumber \\
& + & n_{\rm out} \sum_{\vec q}| V_{\rm out}(\vec k - \vec
 q)|^2 \hat \tau_3 G_q (i\tilde{\omega}_n)\hat \tau_3
\eeq
where $V_{\rm in}$ and $V_{\rm out}$ represent
in-plane and out-of-plane impurity potentials while  
$n_{\rm in}$ and $n_{\rm out}$ their concentrations.
After Kee \cite{Kee} we assume the out of plane impurity 
potential to be of short range type $V(\vec r) = u_0/(r^2 + d^2)$, 
where $d$ is the distance of the impurity from conducting plane
and $u_0$ its scattering amplitude. Fourier transform of this
potential is expressed in terms of modified Bessel function 
of the second kind $K_0(d|\vec k -\vec k'|)$, which is 
strongly decreasing function of its argument. Therefore we 
approximate it by constant $K_0$ for $d|\vec k - \vec k'|> 1$ and zero
otherwise. 
This means that the momentum transfer $|\vec k - \vec k'|$
in the scattering process by the out-of-plane impurities 
is limited to small 
values. In two-dimensions this translates into small angle 
scattering. There is no such limitation on the momentum 
transfer in the scattering by the 
in-plane impurities. Therefore we take \cite{Kee,Nak}
$$
 V_{\rm out}(\vec q) = \left\{\begin{array}{ll}
 V_{\rm out} & {\rm for}~~~~\phi - \phi' < \theta_c\\ 0 & {\rm otherwise}
\end{array}\right.
$$
and
$$
 V_{\rm in}(\vec q) = V_{\rm in}.
$$
In the following  we shall calculate the changes of
$T_c$ and $\alpha$ due to both kinds of impurities in
superconductors.

\begin{figure}[h]
\centerline{\label{fig_1} \epsfig{file=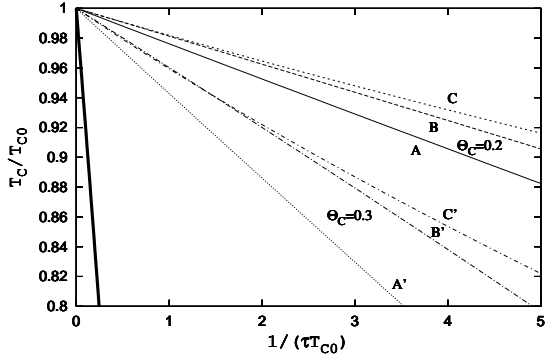,width=10 cm  }}
%\centerline{\epsfig{file=rys2.eps,width=10 cm   }}
\caption{\label{fig1} Effect of in-plane (thick solid line) and
out-of-plane 
impurities (thin lines) in d-wave HTS on $T_{c}$ for 
different values of $\theta_c$  and for  
${1\over 2\tau_{\rm in}}=0$.  The two groups of  curves   correspond 
to two values of $\theta_c =0.2,0.3$, while different labels refer to diferent 
angle dependent density of states functions:  A and A' are calculated with 
$N(\phi)=1$, B and B' with 
$N(\phi)={\pi \over 2}|\cos 2\phi|$,
C and C' with $N(\phi)={\pi \over 2}|\cos 4\phi|$. Note much weaker influence
of out-of-plane impurities on $T_c$.}
\end{figure}
\subsection{Impurities in d-wave superconductor}

Now we specialise the calculations to specific symmetries of the
order parameter starting  with $d$-wave one with the order parameter
$\Delta(\vec k) = \Delta_0\cos2\phi$. To proceed we introduce 
angle dependent density of states (DOS) function  $N_F(\phi$), normalised 
to its average value, and assume it to be energy independent 
in the energy window near the Fermi energy. The integrals 
over $(\varepsilon - \mu)$ can be easily performed and we find
\beq
&&\hat\Sigma(\vec k,i\tilde\omega_n) = -n_{\rm in} V^2_{\rm in}
  \int^{2\pi}_0
{d\phi'\over 2 } N_F(\phi') {i\tilde\omega_n -
\tilde\Delta(\phi')\hat\tau_1 \over \sqrt{\tilde\omega^2_n +
\tilde\Delta^2(\phi')}} \nonumber\\
&&- n_{\rm out} V^2_{\rm out}   \int^{\phi+\theta_c}_{\phi}
{d\phi'\over 2} N_F(\phi') {i\tilde\omega_n -
\tilde\Delta(\phi')\hat\tau_1 \over \sqrt{\tilde\omega^2_n +
\tilde\Delta^2(\phi')}}\,.
\eeq
We are interested in the change of the superconducting transition
temperature and simplify the equations by neglecting powers 
of $\tilde\Delta(\phi)$ higher than the first.

Even though it is possible to continue calculation for general
$N_F(\phi)$, let us for a moment   take 
$N_F(\phi) = N_F$ independent of $\phi$, as is appropriate 
for the circular Fermi surface, to underline the effect of
small angle scattering only. Assuming that 
$u_n = {\tilde\Delta\over |\tilde\omega_n|}$ is independent 
of the angle $\phi$ we get
\beq
&&\tilde\omega_n - \omega_n = \left({1\over 2\tau_{\rm in}} +
{\theta_c\over 2\tau_{\rm out}}\right){\rm
sign}\tilde\omega_n \, \nonumber\\
&&(\tilde\Delta - \Delta_0) \cos2\phi = {\tilde\Delta\over|\tilde\omega_n|}
\Bigg({1\over \tau_{\rm in}} \int^{2\pi}_0 {d\phi'} \cos
2\phi' \nonumber\\
&&+{1\over 2\tau_{\rm out}} \int^{\phi+\theta_c}_\phi d\phi' 
\cos2\phi'\Bigg) 
\label{renorm}
\eeq
In writing this equation we made use of the fact that
 $\tilde\Delta(\phi)=\tilde\Delta\cos 2\phi$ and $\tilde\Delta$ 
 does not depend on $\phi$ for the considered DOS. Vanishing of
the integral multiplying ${ 1/\tau_{\rm in}}$ in  equation 
  (\ref{renorm}) for $\tilde\Delta$ means that 
the in-plane impurities are strong  
pair-breakers in d-wave supeconductors. 
By the same token non-zero 
 value of the  integral multiplying $1/\tau_{out}$ means 
that the out of plane impurities are much weaker pair breakers. 
This explains \cite{Kee} the long standing puzzle while 
dopant impurities do not kill the d-wave superconductivity in 
HTS.

Performing the integral over $\phi'$ in the last equation 
 and projecting the result onto $\cos 2\phi$ function we get 
 equation
\be
\tilde\Delta(\theta_c) - \Delta_0 = {\tilde\Delta(\theta_c)\over
|\tilde\omega_n|} {1\over 2\tau_{\rm out}} 
\left({\sin 2\theta_c\over 2}-tg2\phi sin^2\theta_c\right)\,,
\ee
\noindent which together with that for $\tilde \omega_n$ allows us to 
solve for 
$u_n(\theta_c) = {\tilde\Delta(\theta_c)\over |\tilde \omega_n|}$. 
The result reads
\begin{widetext}
\be
u_n = {\Delta_0\over \left| \omega_n + \left({\theta_c\over 2\tau_{\rm out}} +
{1\over 2\tau_{\rm in}}\right) \, {\rm sign} \omega_n \right| - {1\over
2\tau_{\rm out}} \left({\sin 2\theta_c\over 2}-tg 2 \phi sin^2 \theta_c \right)}
\ee
\end{widetext}
Gap equation (\ref{gap}) is linearised near $T_c$ and standard
manipulations allow us to find the $T_c$ changes due to simultaneous 
presence of both in-plane and out-of-plane impurities
\be
\ln {T_c\over T_{c_0}} = -\gamma(\theta_c) \sum^\infty_{n=0} {1 \over \left(n +
{1\over 2} \right) \left(n + {1\over 2} + \gamma(\theta_c)\right)}\,,
\label{Tcg}
\ee
where to lowest non-vanishing order in $\theta_c$
$$
\gamma(\theta_c) = {1\over 2\tau_{\rm in} \pi k_B T_c} +
 {1\over 2\tau_{\rm out} \pi k_B T_c}
\left({2\theta_c^3\over 3}\right)
$$
plays a role of pair-breaking parameter. 
Note, that the first and
second order (in $\theta_c$) contributions due to out of plane
impurities vanish. This has previously been derived in \cite{Kee}, 
where the change of the $T_c$ due to the  out of plane impurities has 
been calculated.  The equation (\ref{Tcg}) 
is valid to all orders in ${1/\tau_{\rm in}}$
and  ${1/\tau_{\rm out}}$ but to the lowest non-vanishing
 order in $\theta_c$. It slightly 
generalizes the previous result \cite{Kee} to simultaneous presence of
in- and out-of-plane impurities. 
\begin{figure}[h]
\centerline{\label{fig_2} \epsfig{file=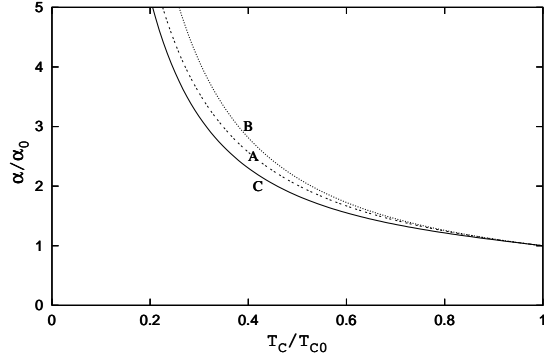,width=10 cm  }}
%\centerline{\epsfig{file=rys2.eps,width=10 cm   }}
\caption{\label{fig2} Effect of the density of states on 
 relative isotope effect. Curve labelled  A refers to constant 
 density of states $N(\phi)=1$, 
 B to  $N(\phi)={\pi \over 2}|\cos 2\phi|$ and 
C to $N(\phi)={\pi \over 2}|\cos 4\phi|$}
\end{figure}
The sum on rhs of equation (\ref{Tcg}) is expressed {\it via} digamma
function $\psi(z)$ as
\be
\ln {T_c\over T_{c_0}} = \psi \left({1\over 2}\right) -
\psi\left({1\over 2} + \gamma(\theta_c)\right)\,.
\label{eq10}
\ee
Figure (\ref{fig1}) shows large  changes of superconducting transition
temperature due to in-plane (thick solid line) 
and much weaker for out-of-plane impurities 
for two values of $\theta_c$ (thin lines).  

 For small values of $\theta_c$ even strong impurities have negligible effect
 on $T_c$.
The change of the isotope shift exponent can be easily calculated 
and found to read
\be
{\alpha_0\over \alpha} = 1 - \gamma(\theta_c) \psi' (\left({1\over 2}
+\gamma(\theta_c)\right)\,.
\label{alpha_d}
\ee
 where $\psi'$ denotes the derivative of the di-gamma function.
 \begin{figure}[h]
\centerline{\label{fig_3} \epsfig{file=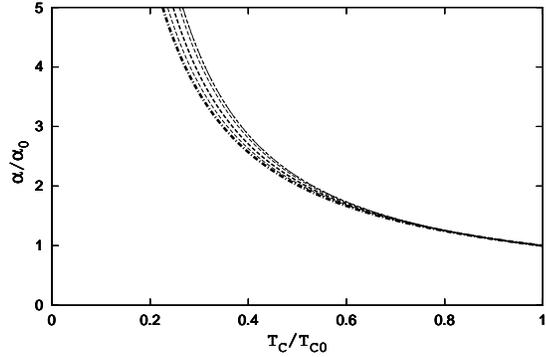,width=10 cm  }}
%\centerline{\epsfig{file=rys2.eps,width=10 cm   }}
\caption{\label{fig3} Simultaneous effect of in-plane  and
out-of-plane 
impurities on the isotope coefficient for various values of {\it x} (from the 
bottom curve x=0.0, 0.01, 0.03, 0.1, 1.0)
 which measure the relative contribution 
 of in-plane impurities to the total 
 scattering rate ({\it c.f} eq.(\ref{tau_x})).}
\end{figure}
This   solution of the impurity problem has been obtained under the
 assumption of constant ({\it i.e} angle independent) density of states 
 function $N_F(\phi)$ and
 scattering rate.
 \begin{figure}[h]
\centerline{ \label{fig_4}\epsfig{file=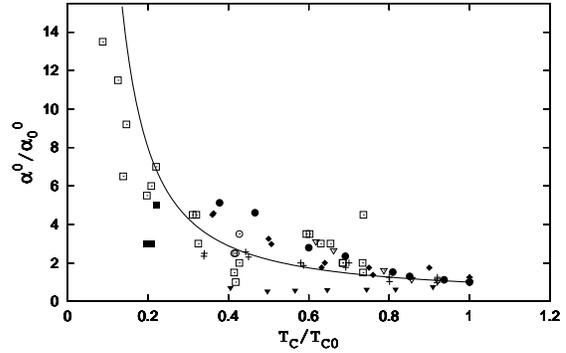, width=10.0cm}}
\caption{ \label{fig4} Solid curve shows the universal
dependence of the normalised isotope coefficient {\it versus}
noermalised transition temperature calculated for the d-wave 
impure superconductor with constant $N_F$.
  The data points are 
experimental values for a number of materials. Optimally doped   
$La_{1.85}Sr_{0.15}Cu_{1-x}M_{x}O_{4}$ with M=Co (opened triangles), 
Zn (filled diamonds) and Ni (crosses); overdoped
 $La_{1.80}Sr_{0.20}Cu_{1-x}M_{x}O_{4}$  with M= Ni (filled traingles),
Fe (filled circles)\cite{babush}; $Y_{1-x-y}Pr_xCa_yBa_2Cu_3O_{7-\delta}$ with x=0.2
and for y=0.15 (opened circles), 0.25 (filled squares);
$YBa_2(Cu_{1-z}Zn_z)_3O_{7-\delta}$ (opened aquares) \cite{gygax}.}
\end{figure}
The  solution of equations (\ref{renorm})
with $\phi$ dependent density of states can 
also be easily obtained by assuming that 
$u_n(\phi') = {\tilde\Delta(\phi')\over|\tilde\omega_n|}$ is 
slowly varying function of the angle for angles
$\phi' \in \langle \phi, \phi +\theta_c\rangle$ and 
we obtain
\begin{widetext}
\be
 \ln{T_c\over T_{c_0}} = {1\over a_0} \int^{\pi/4}_0 d\phi N(\phi) \cos^2
    2\phi \left[\psi({1\over2})   
  -  \psi\left({1\over 2} + \rho_c(\phi,\theta_c)\right)\right]
 \label{eq14iz}
\ee
\end{widetext}
with the pair breaking parameter
\be
\rho_c(\phi,\theta_c) = {1\over 2\pi k_B T_c \tau(\phi,\theta_c)}
\ee
and angle dependent relaxation time $\tau(\phi,\theta_c)$
\be
{1\over \tau(\phi,\theta_c)} = {1\over \tau_{\rm in}} + {1\over \tau_{\rm
out}} \int^{\phi+\theta_c}_\phi d\phi' N(\phi')\left(1-{\cos 2\phi' \over 
\cos 2\phi}\right)
\label{taufi} 
\ee
The parameter $a_0$ is given by
\be
 a_0 = \int^{\pi/4}_0 d\phi N(\phi) \cos^2 2\phi\,.
\ee
The first and second terms contributing to $1/\tau(\phi,\theta_c)$
 come from frequency renormalization by the in- and out-of-plane 
 impurities respectively, while the third one  is due to gap 
 renormalisation by out-of-plane impurities. In-plane impurities 
 do not renormalise the gap, as is seen from equation 
 (\ref{renorm}).
 
The effect of the angle dependent density of states on the 
suppression of $T_c$ is also illustrated in the figure (\ref{fig1}).  
To facilitate comparisons we have properly normalized all the densities of
states. The observed dependences
are easy to understand. Strong   changes, with $\phi$, of the
functions appearing in equation (\ref{eq14iz})  diminish 
the integral 
on its {\it rhs} and this results in a weaker decrease of $T_c$.
Physically it means that due to angular dependence of  spectrum 
the phase space is reduced and scatterers are effectively weaker 
pair breakers.

Straightforward differentiation  leads to
the following expression for the isotope coefficient $\alpha$
\be
 {\alpha_0 \over \alpha} = 1- {1\over a_0} \int^{\pi/4}_0 d\phi N(\phi) \cos^2
    2\phi  \rho_c(\phi,\theta_c)
    \psi^\prime \left({1\over 2} + \rho_c(\phi,\theta_c)\right). 
    \label{eq14alfa}
\ee
The effect of angle dependence in $N_F(\phi)$ on isotope coefficient is 
illustrated in figure (\ref{fig2}). It is very small in relatively 
clean samples (large $T_c/T_{c0}$) and increases for smaller $T_c/T_{c0}$.  
In order to see the simultaneous effect of both types of  
impurities on the isotope coefficient we have assumed that fraction {\it x} of
impurities goes into planes while the rest into out-of-plane positions. 
Assuming that both types of impurities can be 
characterised by the same value of relaxation rate parameter {\it i.e.} 
${1\over \tau_{\rm in}}={1\over \tau_{\rm out}}={1\over \tau}$ we
rewrite (\ref{taufi}) as
\be
{1\over \tau(\phi,\theta_c)} = {1\over \tau }\left[ x +  
(1-x) \int^{\phi+\theta_c}_\phi d\phi' N(\phi')\left(1-{\cos 2\phi' \over 
\cos 2\phi}\right)\right]. 
\label{tau_x}
\ee
Figure (\ref{fig3}) illustrates the changes of $\alpha/\alpha_0$
with x. Bottom curve in the figure corresponds to $x=0$ and thus to the 
changes induced by the out-of-plane impurities while the top one corresponds
 to the case with all impurities occupying in-plane positions. The assumption  
 $\tau_{in}=\tau_{out}$ is not a realistic one. In fact to get the whole curve 
 for $x=0$ one has to take unphysically large values of $1/\tau_{out}T_{c0}$. 
 
 Figure  (\ref{fig4}) shows the changes of the normalised isotope 
 coefficient  with disorder 
 characterised by the ratio $T_c/T_{c0}$ 
 (solid curve)  together with experimental data \cite{babush,gygax} on 
 $La_{1.85}Sr_{0.15}Cu_{1-x}M_{x}O_{4}$ (points). Theoretical line has been
 calculated for $d$ wave superconductor with $N_F(\phi)=1$ and angle independent
 scattering rate. As discussed above angle dependence adds a new degree of
 freedom. 

\subsection{Impurity effects in $s$-wave layered superconductor}

In the $s$-wave superconductor the order parameter 
$ \Delta(\vec k) = \Delta_0$,  and nonmagnetic  isotropic 
impurities do change neither $T_c$ nor the isotope 
coefficient $\alpha$. This is true for both in-plane and out-of-plane 
impurities.

Contrary, the magnetic (in-plane) impurities are pair breakers in
s-wave superconductors and do change $T_c$ \cite{abrikgor,Maki}
\be
\ln {T_c\over T_{co}} = \psi \left({1 \over 2}\right)-
\psi \left({1\over 2} + \rho_c \right) ,
\ee
where $\rho_c={1 \over 2\pi k_B T_c\tau_{m}}$, $\tau_m$ is the 
corresponding magnetic relaxation time.  
This leads to the changes of 
the isotope effect
\be
 {\alpha_{0} \over \alpha}=1-\rho_c \psi' \left({1\over 2} + \rho_c \right).
\ee

\subsection{Impurity effects of $p$-wave layered superconductor}

The order parameter of the $p$-wave superconductor is taken
here as
\be
 \Delta(\phi) = \Delta_0  \cos \phi.
\ee
Repeating the  calculations  for the in-plane and out-of-plane 
impurities in p-wave superconductor  with constant density of states
we get the equations 
(\ref{Tcg}-\ref{alpha_d}) with
$$
\gamma(\theta_c) = {1\over 2\tau_{\rm in} \pi k_B T_c} +
 {1\over 2\tau_{\rm out} \pi k_B T_c}
\left(\theta_c^3\over 6 \right).
$$ 
It is important to note that for constant density of states
function the calculated changes of the
isotope coefficient follow the universal curve independently 
what is the cause changing the superconducting transition
temperature. The universal $\alpha/\alpha_0$ {\it vs.} $T_c/T_{c0}$
dependence is presented by the curve A in figure (\ref{fig2}). Note that 
numerically this
dependence for {\it d}-wave superconductor is the same
as that obtained for magnetic impurities in {\it s}-wave material.

\section{Anisotropic magnetic and nonmagnetic impurity potentials}

High temperature superconductors are strongly anisotropic systems 
with relatively low carrier concentration and layered structure. 
This means that the screening is not very effective and the 
impurity - quasiparticle interaction is anisotropic. 
This motivates the study of anisotropic impurities \cite{haran} 
in HTS.
 The effect of anisotropic magnetic
and nonmagnatic impurities on $T_c$ of superconductors with 
the general form of the order parameter 
$\Delta(\vec k)=\Delta e(\vec k)$ have been considered in
Ref. \cite{haran}.
These authors   have  taken the momentum - dependent 
impurity potential 
$u(\vec k, \vec k')= v(\vec k, \vec k') + J(\vec k,
\vec k')\vec S \cdot \vec \sigma$,
(where $\vec S$ is a classical spin of the impurity and
 $\vec\sigma$ the electron
spin density) and assumed a separable form of scattering 
probabilities
\beq
 v^2(\vec k, \vec k')= v_0^2 + v_1^2 f(\vec k) f(\vec k')\\
 J^2(\vec k, \vec k')= J_0^2 + J_1^2 g(\vec k)g(\vec k')\
\eeq
where $v_0(v_1)$, $J_0(J_1)$ are isotropic (anisotropic) 
scattering amplitudes for non-magnetic and magnetic potentials.
$f(\vec k)$, $g(\vec k)$ are the momentum - dependent anisotropy
functions in the nonmagnetic
and magnetic scattering channel, respectively.
The averages
over the Fermi surface of $f(\vec k)$ and $g(\vec k)$ 
vanish $\langle f(\vec k)\rangle_{\rm FS} = 0$ and are
normalised as $\langle f^2(\vec k)\rangle_{\rm FS} = 1$.
The change of transition temperature calculated in the Born approximation
 is given in  \cite{haran}
%\beq
%&&\ln {T_{c} \over T_{co}} = (1 - <e>^{2} - <ef>^{2} )
%  \Bigg[\psi \left({1\over 2} \right)- \nonumber\\
%&&\psi \left({1\over 2}
%   + {\Gamma_{0}+G_{0} \over 2\pi k_B T_{c}} \right) \Bigg]
%+ <ef>^{2} \Bigg[\psi \left({1\over 2} \right)- \nonumber\\
%&&\psi \Bigg({1\over 2} + {\Gamma_{0}+G_{0}
%  +G_{1}-\Gamma_{1} \over 2\pi k_B T_{c}} \Bigg) \Bigg]\nonumber\\
%&&+ <e>^{2} \Bigg[\psi \left({1\over 2} \right)- \psi \Bigg({1\over 2}
%   + {2G_{0}\over 2\pi k_B T_{c}} \Bigg) \Bigg],
%\eeq
and the isotope effect is found to be
\beq
&&{\alpha_{0} \over \alpha} = 1- \left(1 - <e>^{2} - <ef>^{2} \right)
\cdot \nonumber\\
&&\cdot \left( {\Gamma_{0}+G_{0} \over 2\pi k_B T_{c}} \right)\psi'
 \left({1\over 2}
   + \left( {\Gamma_{0}+G_{0} \over 2\pi k_B T_{c}} \right) \right)\nonumber\\
&&- <ef>^{2} \left( {\Gamma_{0}+G_{0}+G_{1}-\Gamma_{1} \over 2\pi k_B T_{c}} 
\right)\cdot \nonumber\\
&& \cdot  \psi' \left({1\over 2} + {\Gamma_{0}+G_{0}
  +G_{1}-\Gamma_{1} \over 2\pi k_B T_{c}} \right)\nonumber\\
&&- <e>^{2} \left( {2G_{0}\over 2\pi k_B T_{c}} \right) \psi' \left({1\over 2}
   + {2G_{0}\over 2\pi k_B T_{c}} \right),
\eeq
where $\Gamma_0=\pi n_i N_0v_0^2$, $\Gamma_1=\pi n_i N_0v_1^2$,
$G_0=\pi n_i N_0 J_0^2 S(S+1)$ and $G_1=\pi n_i N_0 J_1^2 S(S+1)$
are the  respective scattering rates.
\begin{figure}[h]
\centerline{ \label{fig_5}\epsfig{file=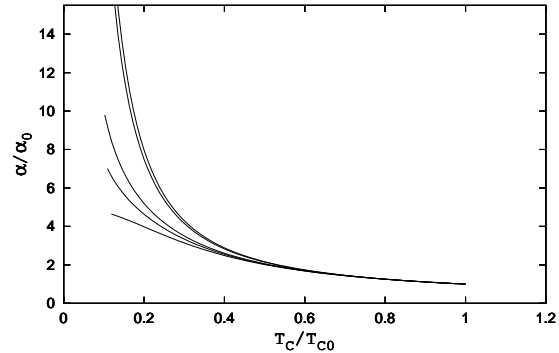,width=10.0cm}}
\caption{ \label{fig5}The isotope coefficient of  
 d-wave supercondutor (${<e>}=0.0$), for different 
values of the normalized anisotropic scattering rate (from top):
${\Gamma}_{1}/{\Gamma}_{0}=0.0, 0.5, 0.9, 0.95, 1.0$,
 We have taken $<ef>^2=0.2$ and assumed non-magnetic impurities.}
\end{figure}
The dependence of $\alpha/\alpha_0$ on $T_c/T_{c0}$ has been shown in
Fig.(\ref{fig5}). The value of anisotropy of the $<ef>^2$ and the 
scattering $\Gamma_1$ present new degrees of freedom which can be
used to fit experimental data. The universality seen for
isotropic scattering is lost in these scenario.

\section{Comparison with experiments}

Let us start the comparison of the obtained results with existing
experimental data with a word of caution. There may exist a number 
of factors which can affect the value of isotope coefficient of 
the impure system $\alpha$. In particular electron-phonon 
interaction and the phonon spectrum may change after the impurity doping.  
We do not take such effects into account.

 We have also implicitely assumed that electron-phonon interaction does 
 play a role in driving the superconducting instability of the 
 system and makes the clean material coefficient $\alpha_0$ non-zero olbeit it
 may take on very small value.
  This, however, seem to be well established by  
 various experimental techniques \cite{Mul}, \cite{Con}.

Here we concentrate on the oxygen isotope effect, but it has to be
noted that interesting results have also been obtained in the studies of
copper isotope substitutions.
The negative value of $\alpha_{\rm Cu}$ observed in some underdoped
YBa$_2$Cu$_3$0$_{7-\delta}$ samples \cite{Fra1}, \cite{Fra2} have 
been
recently explained \cite{Abr} as due to scattering of electrons by
low-frequency phonons with large momenta. It is to be checked 
whether the same mechanism is responsible for negative oxygen 
isotope effect in very clean $Sr_2RuO_4$ p-wave superconductor 
\cite{strontium}.

The oxygen isotope shift in high temperature superconductors 
does depend on the concentration of carriers.
The explanation of this dependence has been a subject of number of
papers \cite{Szcze}-\cite{Gre}. The authors invoke such effects as:
anharmonicity,
zero point motion, mass dependence of carrier concentration, energy
dependence of the density of states, opening of the normal state
pseudo-gap and the $T_c$ changes due to impurities.

In the present work we contribute to the elucidation of the role 
played by various impurities in changing $T_c$ and $\alpha$ of
superconductors. Because the $T_c$ reduction due to 
 dopant impurities like $Sr$ in La$_{2-x}$Sr$_x$CuO$_4$
is small their influence 
 on $\alpha$ follows the standard curve like one 
presented in the figure (\ref{fig4}). The departures would be seen for 
smaller ratio of $T_c/T_{c0}$ as is evident from figure (\ref{fig3}).
 Small values of $\theta_c$ 
which reflect weaker pair breaking character of out-of-plane impurities 
does not influence the slope of the $\alpha/\alpha_0$
$vs.$ $T_c/T_{c0}$ curve. Isotope changes due
to impurities in systems with the same bare scattering rate
but with differing values of $\theta_c$ follow the same universal curve.

On the other hand the effect of anisotropic impurity scattering 
changes the universal dependence. Isotope coefficients
of the systems with various degree of 
anisotropy  follow slightly differing curves. 
This additional degree of freedom allows for a better
fit to the experimental data. The fit, however is ambiguous. The same 
changes can be induced by anisotropy of scattering 
and the anisotropy of the order parameter.
 
\section{Isotope effect in impure striped cuprates}

The strong sensitivity of $T_c$ to in-plane impurities in high temperature
superconductors, 
independently on their magnetic nature, and the difficulties of 
the existing theories to fully describe the wealth of 
experimental data has resulted in  new approaches to the problem. 
In a recent work Morais Smith and coworkers \cite{CMS} have 
developed the scenario of $T_c$ suppression by Zn impurities, 
based on the stripe picture of high $T_c$ oxides. Assuming that 
the doped Zn does not affect the superfluid density the authors 
considered the increase of local inertia of the stripe due to 
pinning forces and local slowing down of their dynamics. The 
calculations within this model give , in agreement with experimental
findings,
the suppression of $T_c$ which is linear in the 
concentration of impurities $z$. It also gives 
the quadratic dependence of 
critical Zn concentration $z_c$ ($z_c$ is concentration at
which superconductivity disappears) on the superconducting 
transition temperature of Zn free ('clean') systems $T_{c_0}$.

The resulting formula for $T_c$ suppression in the regime of
incompressible stripes is \cite{CMS}
\be
 {T_c\over T_{c_0}} = 1 - {\gamma z\over T_{c_0}^2}\,,
\ee
where $z$ is concentration of Zn (or other in-plane impurities), 
$\gamma$ is a factor depending {\it inter alia} on the stripe 
distance and lattice constant $a$. In the optimally doped and 
overdoped region, where charged stripes behave as an 
compressible quantum fluid the $T_c$ reduction has been found 
to be universal and given by
\be
 {T_c\over T_{c_0}} = 1 - {z\over z_c}\,.
\ee
with constant $z_c$.

In the stripe senario we find that the change of isotope coefficient due to
impurities reads
\be
 \alpha/\alpha_0 = {2 \over \left({T_c/T_{c_0}}\right)} - 1
 \label{stripe1}
\ee
in the incompressible region and 
\be
\alpha/\alpha_0 = 1
\label{stripe2}
\ee 
 in the
compressible region.

Roughly inverse dependence of $\alpha/\alpha_0$ on $T_c/T_{c_0}$ 
at low $T_c$ is in good qualitative agreement with experimental 
data (cf. figure (\ref{fig4})) and gives
 credit to the stripe picture of superconductivity   
 in this material.

It is, however, hard to explain the difference between Fe and Ni 
substitution to the otherwise overdoped $La_{1.85}Sr_{0.15}CuO_4$ 
sample. According to the theory \cite{CMS} 
 one expects in this material incomensurate stripes and thus  no 
influence of impurities on both $T_c$ and $\alpha$. 
The difficulty arises from the fact that  
both impurities ($Ni$ and $Fe$) do change 
the transition temperature but the isotope 
shift is nearly constant for $Ni$ substitution. 
One explanation is that  
concentration of carriers may change in the doping process. 
The divalent $Ni$ does not change the
concentration and the position of the Fermi level remains roughly constant 
while
trivalent Fe ions change it driving the system effectively 
into underdoped or
optimally doped regime. If this is true the $Fe$ doped 
system is thus expected,
within stripe scenario, to change both $T_c$ and $\alpha$.  
\begin{figure}[h]
\centerline{ \label{fig_6}\epsfig{file=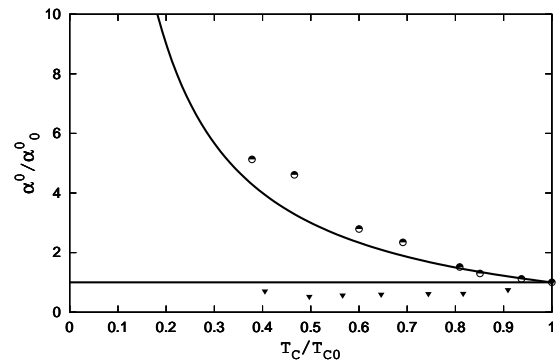, width=10.0cm}}
\caption{ \label{fig6} The isotope effect in impure striped 
cuprates: the continuous line is given by Eq. (\ref{stripe1},\ref{stripe2}). 
The data points are 
experimental values for overdoped   
$La_{1.80}Sr_{0.20}Cu_{1-x}M_{x}O_{4}$, with Ni (filled triangles), 
Fe (circles) \cite{babush}.}
\end{figure}
It is obvious that the  isotope coefficient $\alpha$ of impure 
system can  take non-zero values only for those systems for which
$T_{c0}$ depends on isotope mass and $\alpha_0 \neq 0$. 
The  question thus arises whether the formation of stripes and 
 the driving mechanism of superconductivity in striped metal 
 are of purely electronic origin or electron-phonon interaction 
 plays the role to make $\alpha_0 \neq 0$. 
 These issues have been recently addressed 
 \cite{Lanz}. The XANES experiments revealed large oxygen isotope
  effect on the stripe formation temperature $T^\star$, as also
on effective supercarrier mass. This shows that electron phonon 
interaction does play a role in a stripe model of 
superconductivity and validates the above analysis and 
conclusions.

\section{Conclusions}
We have studied the effect of various impurities 
on the isotope coefficient $\alpha$.
While the in-plane and out-of-plane impurities affect
the superconducting transition temperature
quite differently ({\it c.f.} Fig. 1),
their influence on $\alpha$
is universal for   angle independent
  scattering relaxation rate and  density of states function. 
   
Experimental changes of $\alpha/\alpha_0$  with $T_c/T_{c0}$
for a number of high temperature copper oxides 
 can be well described solely by the effect of impurities 
 on $d$-wave superconductors. 
The angle dependence of relaxation rate and $N_F(\phi)$ or  the
simultaneous presence of anisotropic scatterers of   
magnetic and non-magnetic nature 
adds new degrees of freedom which can be used to quantitatively
describe the data. 

These mechanisms of $T_c$ and $\alpha$ changes 
operate for both dopant (like $Sr$) and 
extra  impurities (like  $Ni$ or $Zn$). However, they do not differentiate
between underdoped and overdoped systems. On the other
hand the stripe model of superconductivity describes
the detrimental effect of $Ni$ or $Zn$ on $T_c$. 
The stripe theory predicts 
increase of $\alpha$  with impurity concentration
on  the underdoped side of the phase diagram and
 no change of it in the overdoped region. This agrees
with data  \cite{babush} 
on $Ni$ and $Fe$ doped  $La_{1.80}Sr_{0.20}Cu_{1-x}Ni(Fe)_{x}O_{4}$.

\noindent
{\bf Acknowledgements:} This work has been supported by KBN
grant no. 2P03B 106 18.

\end {document}